\documentclass[11pt]{article}
\newcommand{\be}{\begin{equation}}
\newcommand{\ee}{\end{equation}}
\newcommand{\bea}{\begin{eqnarray}}
\newcommand{\eea}{\end{eqnarray}}
\newcommand{\sn}{{\rm sn}}

\newcommand{\dn}{{\rm dn}}
\newcommand{\cn}{{\rm cn}}
\newcommand{\sech}{{\rm sech}}

\begin{document}

\vspace{0.5in}
\begin{center}
{\LARGE{\bf Linear Superposition of Quadratic Functions in a Fifth Order 
KdV-Type Equation}}
\end{center}
\vspace{0.3in}
\begin{center}
{\LARGE{\bf Avinash Khare}} \\
{Physics Department, Savitribai Phule Pune University \\
 Pune 411007, India}
\end{center}

\begin{center}
{\LARGE{\bf Avadh Saxena}} \\ 
{Theoretical Division and Center for Nonlinear Studies, 
Los Alamos National Laboratory, Los Alamos, New Mexico 87545, USA}
\end{center}

\vspace{0.3in}
\noindent{\bf {Abstract:}}

We show that a fifth order KdV-type equation admits several real as well as
complex parity-time reversal or PT-invariant solutions with linear superposition 
of quadratic functions involving Jacobi elliptic functions of the form $\dn^2(x,m)$, 
$\cn(x,m)\dn(x,m)$, $\sn(x,m) \cn(x,m)$ and $\sn(x,m)\dn(x,m)$. These 
results must be contrasted with only partial superposition of such functions in 
Korteweg-de Vries (KdV), $\phi^3$ and a few other nonlinear equations.

\section{Introduction}

One of the hallmarks of a linear system is the superposition principle. In 
particular, if $f(x)$ and $g(x)$ are two solutions of a linear equation, then
any arbitrary superposition of these two functions is also a solution of this
equation. Unfortunately, because of the nonlinear term, this superposition
principle does not hold in nonlinear systems. In particular, even if $f(x)$ and 
$g(x)$ are two solutions of a nonlinear equation, their linear superposition
is in general not a solution of that nonlinear equation. However, in recent 
years it has been shown that several nonlinear equations \cite{sukhatme}, both 
integrable and nonintegrable, discrete and continuous, as well as local and nonlocal 
admit linear superposition. In particular, in several nonlinear models it has been 
shown that not only the Jacobi elliptic {\it linear} functions $\dn(x,m), \cn(x,m)$ and 
$\sn(x,m)$ are solutions of these nonlinear equations but even linear superpositions  
of these functions, i.e., $A\dn(x,m)+B\cn(x,m), A\dn(x,m) + B\sn(x,m)$ and 
$A\cn(x,m)+B\sn(x,m)$ are also the solutions of the same nonlinear equation 
for a fixed value of $B/A$ \cite{ks1,ks2,ks3}. Here, $0 \le m \le 1$ is the 
modulus of the Jacobi elliptic functions \cite{as}. 

However, to the best of our knowledge, no example is known in the literature 
where a nonlinear equation admits a linear superposition of {\it quadratic} functions.  
What is known so far is only the ``partial superposition'' of quadratic functions in a 
few nonlinear equations such as KdV, $\phi^3$, etc. \cite{ks1,ks2,ks3}.
In particular, while $\dn^2(x,m)$ is a solution of these nonlinear equations, neither 
$\cn(x,m) \dn(x,m)$, $\sn(x,m) \dn(x,m)$ nor $\sn(x,m) \cn(x,m)$ are the solutions 
of these nonlinear equations. Nevertheless, a superposition of $\dn^2(x,m)$ with any 
of these three quadratic functions, i.e. $A\dn^2(x,m)+B\cn(x,m)\dn(x,m), 
%\newline
A\dn^2(x,m)+B\sn(x,m)\cn(x,m)$ as well as $A\dn^2(x,m)+B\sn(x,m)\dn(x,m)$  
are the solutions of these nonlinear equations for a fixed ratio of $B/A$. 
Similarly, while $\sech^2(x)$ is a solution of these nonlinear equations, 
$\sech(x)\tanh(x)$ is not a solution of these nonlinear equations. However, a 
superposition of the two, i.e. $A\sech^2(x)+B\sech(x)\tanh(x)$ is still a 
solution of these nonlinear equations for a fixed ratio of $B/A$.

The purpose of this paper is to discuss an example where not only two quadratic
functions, say $f(x)$ and $g(x)$, are solutions of the nonlinear equation but 
even a linear superposition $Af(x)+Bg(x)$, for fixed value of $B/A$, is a 
solution of the same nonlinear equation. In particular, we consider the 5th 
order KdV-type equation which was discussed extensively by Karpman 
\cite{kar,kar1,kar2,kar3} and others \cite{dkn,bis,waz}
\be\label{1}
u_t + \alpha u^2(x,t) u_x + \beta u_{xxx} + \gamma u_{xxxxx} = 0\,.
\ee 
On using the moving ansatz
\be\label{2}
y = \eta(x-vt)\,,
\ee
in Eq. (\ref{1}), it takes the form
\be\label{3}
-v u_y +\alpha u^2(y) u_y + \beta \eta^2 u_{yyy} + \gamma u_{yyyyy} = 0\,.
\ee
We show that Eq. (\ref{3}) admits 10 periodic and two hyperbolic quadratically 
superposed solutions. In particular, we show that Eq. (\ref{3}) not only
admits $\dn^2(y,m), \cn(y,m) \dn(y,m), \cn(y,m) \sn(y,m)$ and 
$\sn(y,m) \dn(y,m)$ as solutions but even admits all possible superpositions 
of these functions as a solution of the same nonlinear equation, Eq. (\ref{3}). 
In the hyperbolic limit of $m = 1$, Eq. (\ref{3}) not only admits both $\sech^2(x)$ 
and $\sech(x) \tanh(x)$ as solutions but even their complex (parity-time reversal or) 
PT-invariant superposition with PT-eigenvalue of $+1$ as well as $-1$ are the 
solutions of Eq. (\ref{3}). Specifically, out of the 10 periodic solutions while two of 
them are real solutions, 8 solutions are complex PT-invariant solutions with 4 having 
PT-eigenvalue $+1$ and another 4 having PT-eigenvalue $-1$. These solutions 
are of the form $Af(x)+Bg(x)$ for a fixed ratio $B/A$. 

\section{Six Solutions of the KdV-like Equation} 

Let us first show that $\dn^2(y,m)$, $\cn(y,m) \dn(y,m)$, $\cn(y,m)\sn(y,m)$, 
$\sn(y,m)\dn(y,m)$ and hence $\sech^2(y)$ and $\sech(y) \tanh(y)$ are the 
solutions of Eq. (\ref{3}). \\ 

{\bf Solution I}

It is not difficult to show that Eq. (\ref{3}) admits the solution
\be\label{5}
u(y) = A \dn^2(y,m)\,,
\ee
provided
\be\label{6}
\alpha A^2 = -360 \gamma \eta^4\,,~~\beta \eta^2 = -20(2-m) \gamma \eta^4\,,
~~v = -8[8 m^2+23(1-m)]\gamma \eta^4\,.
\ee

{\bf Solution II}

Eq. (\ref{3}) also admits the solution
\be\label{7}
u(y) = A \sqrt{m} \cn(y,m) dn(y,m)\,,
\ee
provided
\be\label{8}
\alpha A^2 = -360 \gamma \eta^4\,,~~\beta \eta^2 = -10(1+m)\gamma \eta^4\,,~~
v = -[86m -11(1+m^2)]\gamma \eta^4\,.
\ee

{\bf Solution III}

In the limit $m = 1$, the solutions I and II go over to the hyperbolic 
solution
\be\label{9}
u(y) = A \sech^2(y)\,,
\ee
provided
\be\label{10}
\alpha A^2 = -360 \gamma \eta^4\,,~~\beta \eta^2 = -20\gamma \eta^4\,,~~
v = - 64 \gamma \eta^4\,.
\ee
Notice that for the solutions I, II and III, the sign of $\alpha, \beta, v$ is 
the same and is opposite to that of $\gamma$.

{\bf Solution IV}

Eq. (\ref{3}) also admits the solution
\be\label{11}
u(y) = A \sqrt{m} \sn(y,m) dn(y,m)\,,
\ee
provided
\be\label{12}
\alpha A^2 = 360 \gamma \eta^4\,,~~\beta \eta^2 = 10(2m-1)\,,~~
v = [11+64m(1-m)]\gamma \eta^4\,.
\ee

{\bf Solution V}

Eq. (\ref{3}) also admits the solution
\be\label{13}
u(y) = A m \sn(y,m) cn(y,m)\,,
\ee
provided
\be\label{14}
\alpha A^2 = 360 \gamma \eta^4\,,~~\beta \eta^2 = 10(2-m)\,,~~
v = [11-64(1-m)]\gamma \eta^4\,.
\ee

{\bf Solution VI}

In the limit $m = 1$, the solutions IV and V go over to the hyperbolic 
solution
\be\label{15}
u(y) = A \sech(y) \tanh(y)\,,
\ee
provided
\be\label{16}
\alpha A^2 = 360 \gamma \eta^4\,,~~\beta \eta^2 = 10 \gamma \eta^4\,,~~
v = 11 \gamma \eta^4\,.
\ee
Notice that for the solutions V, VI and IV (in case $m > 1/2$), $\alpha, \beta, \gamma,
v$ have the same sign, i.e. either all are positive or all are negative. 

We now show that Eq. (\ref{3}) not only satisfies these 6 solutions but also 
admits 10 periodic quadratic superposed solutions and two hyperbolic quadratically 
superposed solutions. We first present two real quadratically superposed 
periodic solutions and later present 8 complex PT-invariant periodic solutions 
and finally the two hyperbolic quadratically superposed complex PT-invariant 
solutions. \\ 

\section{Two Real Quadratically Superposed Periodic Solutions} 

{\bf Solution VII}

Remarkably, not only $\dn^2(y,m)$ and $\cn(y,m) \dn(y,m)$ but even their real superposition
\be\label{17}
u(y) = [A \dn^2(y,m) + B \sqrt{m} \cn(y,m) \dn(y,m)]\,,~~0 < m < 1\,,
\ee
is an exact solution of Eq. (\ref{3}) provided
\be\label{18}
B = \pm A\,,~~\alpha A^2 = -90 \gamma \eta^4\,,~~\beta \eta^2 = -5(5-m)\gamma
\eta^4\,,~~v = -4[16-m(1-m)]\gamma \eta^4\,.
\ee

{\bf Solution VIII}

Not only $\sn(y,m) \dn(y,m)$ and $\sn(y,m) \cn(y,m)$ but even their real 
superposition
\be\label{19}
u(y) = [A \sqrt{m} \sn(y,m) \dn(y,m) + B m \sn(y,m) \cn(y,m)]\,,~~0 < m < 1\,,
\ee
is an exact solution of Eq. (\ref{3}) provided
\be\label{20}
B = \pm A\,,~~\alpha A^2 = 90 \gamma \eta^4\,,~~\beta \eta^2 = 5(1+m)\gamma
\eta^4\,,~~v = [19m -4(1+m^2)]\gamma \eta^4\,.
\ee

\section{Complex PT-invariant Quadratically Superposed Periodic Solutions}

We now present 8 complex PT-invariant quadratically superposed periodic 
solutions and the corresponding two quadratically superposed complex 
PT-invariant hyperbolic solutions.

{\bf Solution IX}

Not only $\dn^2(y,m)$ and $\sn(y,m) \cn(y,m)$ but even their complex PT-invariant 
superposition
\be\label{21}
u(y) = [A \dn^2(y,m) + iB m \cn(y,m) \sn(y,m)]\,,
\ee
is an exact solution of Eq. (\ref{3}) provided
\bea\label{22}
&&B = \pm A\,,~~\alpha A^2 = -90 \gamma \eta^4\,,~~\beta \eta^2 = -5(2-m)\gamma
\eta^4\,, \nonumber \\
&&v = -[4m^2 +34(1-m)]\gamma \eta^4\,.
\eea 

{\bf Solution X}

Not only $\dn^2(y,m)$ and $\sn(y,m) \dn(y,m)$ but even their complex PT-invariant 
superposition
\be\label{23}
u(y) = [A \dn^2(y,m) + iB \sqrt{m} \dn(y,m) \sn(y,m)]\,,
\ee
is an exact solution of Eq. (\ref{3}) provided
\bea\label{24}
&&B = \pm A\,,~~\alpha A^2 = -90 \gamma \eta^4\,,~~\beta \eta^2 = -5(5-4m)\gamma
\eta^4\,, \nonumber \\
&&v = -[64(1+m^2) -124m]\gamma \eta^4\,.
\eea 

{\bf Solution XI}

Not only $\cn(y,m) \dn(y,m)$ and $\sn(y,m) \cn(y,m)$ but even their complex 
PT-invariant superposition
\be\label{25}
u(y) = [A\sqrt{m} \cn(y,m) \dn(y,m) + iB m \sn(y,m) \cn(y,m)]\,,
\ee
is an exact solution of Eq. (\ref{3}) provided
\bea\label{26}
&&B = \pm A\,,~~\alpha A^2 = -90 \gamma \eta^4\,,~~\beta \eta^2 = -5(2m-1)\gamma
\eta^4\,, \nonumber \\
&&v = -[4 -11m(1-m)]\gamma \eta^4\,.
\eea 

{\bf Solution XII}

Not only $\cn(y,m) \dn(y,m)$ and $\sn(y,m) \dn(y,m)$ but even their complex 
PT-invariant superposition
\be\label{27}
u(y) = \sqrt{m}[A \cn(y,m) \dn(y,m) + iB \sn(y,m) \dn(y,m)]\,,
\ee
is an exact solution of Eq. (\ref{3}) provided
\bea\label{28}
&&B = \pm A\,,~~\alpha A^2 = -90 \gamma \eta^4\,,~~\beta \eta^2 = -5(2-m)\gamma
\eta^4\,, \nonumber \\
&&v = -[4 m^2-11(1-m)]\gamma \eta^4\,.
\eea 

{\bf Solution XIII}

In the hyperbolic limit of $m = 1$, all four solutions IX to XII go over
to the hyperbolic complex PT-invariant solution with PT-eigenvalue $+1$ 
\be\label{29}
u(y) = [A \sech^2(y) + iB \sech(y) \tanh(y)]\,,
\ee
provided
\be\label{30}
B = \pm A\,,~~\alpha A^2 = -90 \gamma \eta^4\,,~~\beta \eta^2 = -5\gamma
\eta^4\,,~~v = -4 \gamma \eta^4\,.
\ee 

{\bf Solution IXV}

Not only $\dn^2(y,m)$ and $\sn(y,m) \cn(y,m)$ but even their another complex 
PT-invariant superposition, i.e.
\be\label{31}
u(y) = [A m \cn(y,m) \sn(y,m) + iB \dn^2(y,m)]\,,
\ee
is an exact solution of Eq. (\ref{3}) provided
\bea\label{32}
&&B = \pm A\,,~~\alpha A^2 = 90 \gamma \eta^4\,,~~\beta \eta^2 = -5(2-m)\gamma
\eta^4\,, \nonumber \\
&&v = -[4m^2 +34(1-m)]\gamma \eta^4\,.
\eea 
If we compare the solutions IX and IXV, we see that only the relation between 
$\alpha A^2$ and $\gamma \eta^4$ is of opposite sign while the expression for
$v$ and $\beta \eta^2$ in terms of $\gamma \eta^4$ remains unchanged. It turns
out that this is also the case for the solutions X vs XV, XI vs XVI, XII vs XVII and
XIII vs XVIII respectively, and hence we will not repeat this comment after 
the next four solutions. 

{\bf Solution XV}

Not only $\dn^2(y,m)$ and $\sn(y,m) \dn(y,m)$ but even their another complex 
PT-invariant superposition
\be\label{33}
u(y) = [A \sqrt{m} \sn(y,m) \dn(y,m)+iB\dn^2(y,m)]\,,
\ee
is an exact solution of Eq. (\ref{3}) provided
\bea\label{34}
&&B = \pm A\,,~~\alpha A^2 = 90 \gamma \eta^4\,,~~\beta \eta^2 = -5(5-4m)\gamma
\eta^4\,, \nonumber \\
&&v = -[64(1+m^2) -124m]\gamma \eta^4\,.
\eea 
%\newpage
{\bf Solution XVI}

Not only $\cn(y,m) \dn(y,m)$ and $\sn(y,m) \cn(y,m)$ but even their another complex 
PT-invariant superposition
\be\label{35}
u(y) = [A m \sn(y,m) \cn(y,m) + iB \sqrt{m} \cn(y,m) \dn(y,m)]\,,
\ee
is an exact solution of Eq. (\ref{3}) provided
\bea\label{36}
&&B = \pm A\,,~~\alpha A^2 = 90 \gamma \eta^4\,,~~\beta \eta^2 
= -5(2m-1)\gamma \eta^4\,, \nonumber \\
&&v = -[4 -11m(1-m)]\gamma \eta^4\,.
\eea 

{\bf Solution XVII}

Not only $\cn(y,m) \dn(y,m)$ and $\sn(y,m) \dn(y,m)$ but even their another complex 
PT-invariant superposition
\be\label{37}
u(y) = \sqrt{m}[A \sn(y,m) \dn(y,m) + iB \cn(y,m) \dn(y,m)]\,,
\ee
is an exact solution of Eq. (\ref{3}) provided
\bea\label{38}
&&B = \pm A\,,~~\alpha A^2 = 90 \gamma \eta^4\,,~~\beta \eta^2 = -5(2-m)\gamma
\eta^4\,, \nonumber \\
&&v = -[4 m^2-11(1-m)]\gamma \eta^4\,.
\eea 

{\bf Solution XVIII}

In the hyperbolic limit of $m = 1$, all four solutions XIV to XVII go 
over to the hyperbolic complex PT-invariant solution with PT-eigenvalue $-1$ 
\be\label{39}
u(y) = [A \sech(y) \tanh(y) + iB \sech^2(y)]\,,
\ee
provided
\be\label{40}
B = \pm A\,,~~\alpha A^2 = 90 \gamma \eta^4\,,~~\beta \eta^2 = -5\gamma
\eta^4\,,~~v = -4 \gamma \eta^4\,.
\ee 

\section{Conclusion} 
Summarizing, in this paper for the first time one has presented 12 examples of a 
nonlinear model (a 5th order KdV-type equation) admitting linear superposition of 
quadratic functions as solutions. Out of these, 10 are the real and complex 
PT-invariant periodic solutions while the remaining two are examples of complex 
PT-invariant hyperbolic solutions.  This paper raises several open questions, some 
of which are: 

\begin{enumerate}
	
\item In this paper we have presented 5th order KdV-type equation admitting 
linear superposition of quadratic functions as solutions. One obvious question
is: are there other nonlinear models both integrable and nonintegrable, discrete
and continuous, local and nonlocal admitting linear superposition of quadratic
functions as solutions? Note that the 5th order KdV-type model we have discussed 
is a local, nonintegrable nonlinear equation.

\item Perhaps even more interesting, are there nonlinear models, local or 
nonlocal, discrete or continuous, integrable or nonintegrable which admit linear 
superposition of cubic functions as solutions? And more generally, are there
nonlinear models admitting linear superposition of quartic, quintic or more
generally nth order functions as solutions of a nonlinear equation? 

\item Are the 12 quadratically superposed solutions presented in this paper 
stable? Hopefully, some of these, specially the two hyperbolic solutions may be
stable.  This could be studied variationally or numerically. 

\item It would be very interesting if one could find applications of some of these
quadratically superposed solutions in physical contexts.

\end{enumerate}

{\bf Acknowledgment}

One of us, AK, is grateful to Indian national Science Academy (INSA) for the
award of INSA Honorary Scientist position at Savitribai Phule Pune University.
The work at LANL was carried out under the auspices of the US Department 
of Energy NNSA under Contract No. 89233218CNA000001.

\end{document}